# Convergecast with Unbounded Number of Channels

*Roman* Plotnikov[1,*], *Adil* Erzin[1,2], and *Vyacheslav* Zalyubovskiy[1]

[1]Sobolev Institute of Mathematics, 4 Acad. Koptyug Avenue, Novosibirsk, Russia
[2]Novosibirsk State University, 2 Pirogova Str., Novosibirsk, Russia

**Abstract.** We consider a problem of minimum length scheduling for the conflict-free aggregation convergecast in wireless networks in a case when each element of a network uses its own frequency channel. This problem is equivalent to the well-known NP-hard problem of telephone broadcasting since only the conflicts between the children of the same parent are taken into account. We propose a new integer programming formulation and compare it with the known one by running the CPLEX software package. Based on the results of a numerical experiment, we concluded that our formulation is more preferable in practice to solve the considered problem by CPLEX than the known one. We also propose a novel heuristic algorithm, which is based on a genetic algorithm and a local search metaheuristic. The simulation results demonstrate the high quality of the proposed algorithm compared to the best known approaches.

## 1 Introduction

In wireless sensor networks (WSNs), collecting data from all sensor nodes to a distinguished node, called the sink, is one of the most fundamental problems. Due to the limited transmission range of sensor nodes, which follows, in particular, from the need to minimize a communication energy consumption [1], multi-hop communication over a tree-based routing topology is usually used to gather data. Such a pattern is known as convergecast [2].

Since radio communication is the main source of energy consumption, it is important to minimize the amount of transmitted data. One of the ways to optimize communication overhead for sensor nodes is to merge their own data with the received packets by means of some aggregation function. Aggregation convergecast is possible when data are spatially correlated or the goal is to collect some summarized information (e.g. maximum, mean, etc.) In such a scenario, each sensor node needs to send only one packet during the aggregation session.

Because of its ability to provide time bounds, TDMA-based scheduling algorithms are widely used. In a TDMA scheduling, time is divided in equal-length slots under assumptions that each slot is long enough to send or receive one packet [3]. Minimizing time for the aggregated convergecast in this case is equivalent to minimizing the number of time slots required for all packets to reach the sink.

Another important factor of the convergecast protocol is aggregation latency, which is defined as the required number of time slots of the whole data collection process. The problem of minimization of latency is known in literature as minimum-latency aggregation scheduling (MLAS) [4]. The solution of MLAS typically includes two components: a spanning tree rooted at and directed towards the sink node, and schedule, which assigns a transmitting time slot for each tree link so that (1) every node transmits only after all its children in the tree have, and (2) potentially interfered links are scheduled to transmit in different time slots. The last condition means that the TDMA schedule should be interference free, i.e. no receiving node is within the interference range of the other transmitting node. There are two types of interferences or collisions in wireless networks: primary and secondary. A primary collision occurs when more than one node transmits to the same destination. In tree-based aggregation, it corresponds to the case when two or more children of the same parent send their packets in the same time slot. A secondary collision occurs when a node overhears transmissions intended for other node. Such kind of collision is caused by links in the underlying communication graph, but not in the aggregation tree.

The MLAS problem was proven to be NP-hard [5]. Finding an optimal time slot assignment for a given tree is still NP-hard [6]. Therefore, all existing results in literature are heuristic algorithms for finding approximate solutions. Most of them contain two relatively independent phases: aggregation tree construction followed by link scheduling [4, 5, 7].

In this paper, we mainly focus on the first phase – finding the minimum delay aggregation tree assuming that the proper chosen tree would lead to a good solution. Additionally, in this stage we take into account only conflicts between the children of the same parent, i.e. primary collisions. First, such a model is suitable for multichannel transmissions, where secondary interference can be avoided by assigning different frequencies under the assumption that the number of channels is big enough. Moreover, the solution of such a "relaxed" problem can be used as a lower bound of the aggregation latency for the original MLAS, so a tree with a smaller delay can be considered as a better candidate to produce a shorter schedule.

---

[*] Corresponding author: prv@math.nsc.ru

It is worthwhile to mention that the considered problem is equivalent to the problem of finding the optimal broadcasting tree in a graph, also known as a telephone broadcasting problem, which has been proved to be NP-hard [8]. Most existing algorithms construct an aggregation tree based on the shortest path tree (SPT) or connected dominated set (CDS), but as was shown in [9] an optimal solution could be neither SPT nor CDS based. To overcome this issue, we propose a novel heuristic algorithm, which combines a genetic algorithm performing broad search among various aggregation trees with a local search procedure aimed at the pruning of the currently found tree.

In summary, we provide the following contributions towards a better understanding of the aggregated convergecast problem:
- We present an alternative IP formulation for the MLAS problem in case of the absence of secondary collisions and compare it with previously known model.
- We propose a novel heuristic Genetic Local Search (GLS). In contrast to traditional genetic algorithms, GLS uses an embedded local search procedure to further improve the current feasible solution.
- Through extensive simulation experiments, we demonstrate the quality of the solutions achievable by the GLS algorithm vs. the current state-of-the-art methods.

The rest of the paper is organized as follows: The recent research results are overviewed in Section 2. The mathematical formulation of the problem and the comparative analysis of two IP-formulations are given in Section 3. In Section 4, the new heuristic algorithm is described. Simulation results are presented in Section 5, and the paper is concluded in Section 6.

## 2 Related work

Data aggregation for WSNs has been proposed to improve energy efficiency of sensor nodes and consequently prolong network lifetime [10]. Some surveys considering different aspects of the problem have been published [11, 12].

The MLAS problem was first introduced in [5]. The authors proved that the problem is NP-hard even for unit disk graphs, and proposed a $(\Delta - 1)$-approximation algorithm, where $\Delta$ is the maximum node degree in the network graph. In this algorithm, the Shortest Path Tree is created first, which later is used as an input for the scheduling algorithm. In the nearly constant approximation proposed by Huang et al. [7], the data latency bounded by $23R + \Delta - 18$, where $R$ is the network's radius. The algorithm in [13] aims to minimize the data aggregation time by using a Connected Dominated Set (CDS). Moreover, the authors choose the network topology center as the aggregation tree root instead of sink. This allows them to reduce the upper bound to $16R + \Delta - 14$. Wang et al. designed a Peony-tree-based data aggregation algorithm with latency bound $15R + \Delta - 15$ [14]. Based on the properties of neighboring dominators in CDS, Nguyen et al. have improved the algorithm from [13] and have given a proof of upper bound $12R + \Delta - 12$ for their algorithm [15].

In [16], the authors proved that minimizing the schedule length for an arbitrary network in the presence of multiple frequencies is NP-hard and proposed approximation algorithms with worst-case performance bound for geometric networks. They also showed that finding the minimum number of frequencies required to remove all interfering links in an arbitrary network is NP-hard problem. Pan et al. considered convergecast for low-duty-cycled multi-channels WSNs aimed to find a time slot and frequency channel assignment that can minimize the data aggregation delay [17]. The authors proved NP-completeness of the problem and proposed a heuristic scheme, which contains three consecutives phases: tree formation, slot assignment, and channel assignment.

As mentioned earlier, a relaxed version of the MLAS problem, which takes into account only primary collisions, is equivalent to the broadcast time problem. The NP-hardness of this problem is proved in [8]. In [18] it is shown that the problem remains NP-hard even for 3-regular planar graphs. Polynomial-time algorithms for the exact solution are known only for few special graphs: trees [8], complete graphs [19], and unicyclic graphs [20]. An algorithm based on a combinatorial approach with $O(\log n)$ approximated ratio were presented in [21]. As for heuristics, simulation results suggest that the best results are achieved by the algorithms presented in [22] and [23].

## 3 Problem formulation

We consider a WSN consisting of stationary sensor nodes with one sink. All sensors are homogeneous. We use a protocol interference model [24], which is a graph theoretic approach that assumes correct reception of a message if and only if there is no simultaneous transmission within proximity of the receiver. For simplicity, we assume that the interference range is equal to the transmission range. Then the WSN with sink node $s$ can be represented as a graph $G = (V, E)$, where $V$ denotes all the sensor nodes and $s \in V$. An edge $(u, v) \in E$ if and only if the distance between the nodes $u$ and $v$ is within the transmission range.

The problem considered in this paper is defined as follows. Given a connected undirected graph $G = (V, E)$, $|V| = n$, $|E| = m$ and a sink node $s \in V$, find the minimum length schedule of data aggregation from all the vertices of $V \setminus \{s\}$ to $s$ under the following conditions:
- at the same time slot any vertex can either receive or send a message;
- each vertex can receive at most one message during one time slot;
- each vertex can send a message only once.

Since it is convenient to consider the directed edges (*arcs*) when constructing an aggregation tree, we also introduce a directed graph $G_{or} = (V, A)$ constructed from $G$ by replacing each edge with two oppositely directed arcs and excluding the arcs starting from *s*.

### 3.1 Integer Programming formulations

*3.1.2 IP formulation 1*

Tian et al. [9] proposed an IP formulation for the general problem when the elements use the same channel (frequency) and collisions between the vertices (not only between the children of the same parent) are taken into account. The IP-formulation of the problem with an unbounded number of channels may be obtained from this formulation by excluding the corresponding set of constraints as follows.

Let us consider a directed graph $G_{or}' = (V \cup \{s'\}, A \cup (s,s'))$ which is constructed from $G_{or}$ by adding a fictive node $s'$ and an arc $(s,s')$. Let us introduce the variables $x_{a,t}$ for any $a \in A \cup (s,s')$ and $t \in \{1, ..., n\}$: $x_{a,t} = 1$ if an arc $a$ is scheduled to transmit a packet during the time slot $t$, and $x_{a,t} = 0$ otherwise. Let us also denote the set of all arcs starting from $v \in V$ as $S(v)$ and all arcs ending at $v \in V \cup \{s'\}$ as $D(v)$. Then the problem is the following:

$$\sum_{t=0}^{n} t * x_{(s,s'),t} \to_x \min \quad (1)$$

$$\sum_{a \in S(v)} \sum_{t=0}^{n} x_{a,t} = 1, \forall v \in V \quad (2)$$

$$\sum_{t'=t+1}^{n} x_{a't'} \leq 1 - \sum_{a \in S(v)} x_{a,t},$$
$$\forall v \in V \cup \{s'\} \; \forall a' \in D(v) \; \forall t \quad (3)$$

$$\sum_{a \in S(v)} x_{a,t} + \sum_{a' \in D(v)} x_{a',t} \leq 1 \; \forall v \in V \cup \{s'\} \; \forall t \quad (4)$$

In this formulation the time slot when *s* sends a message to $s'$ is taken as an objective function (1). Constraints (2) guarantee that each vertex can transmit data only once. Constraints (3) ensure that, once a vertex transmits, it can no longer receive messages. Constraints (4) hold the requirement that each vertex can only transmit or receive a message during each time slot. Note that the formulation (1)-(4) contains $O(nm)$ variables and $O(n^2 + nm)$ constraints.

*3.1.1 IP formulation 2*

The solution space of the formulation (1)-(4) is rather large: in the case of dense graph $G$ the number of variables may be close to $O(n^3)$ as well as the number of variables of the dual problem. For the efficiency of branch and bound-based exact methods IP formulations of less size are more preferable. Therefore below we propose another IP formulation with $O(n^2)$ variables and $O(n^3)$ constraints.

Let us number all vertices $V = \{v_0 = s, v_1, ..., v_{n-1}\}$ and introduce the following variables. Let $t_i \in \{1, ..., n-1\}$ be the time slot of data sending by the vertex $v_i \in V$; $u_i$ be the number of edges in the path from $v_i$ to $s$ in the convergecasting tree ($u_0 = 0$); $L$ be the length of a schedule; $x_{ij}$ be equal to 1 if $v_i$ sends a message to $v_j$ and 0 otherwise; $y_{ij}$ is equal to 1 if $t_i \geq t_j$ and 0 otherwise. Then the IP-formulation can be written in the following form:

$$L \to_{x, u, t, y, L} \min \quad (5)$$

$$L \geq t_i, \; i = 1, ..., n \quad (6)$$

$$\sum_{j=0}^{n} x_{ij} = 1, \; i = 1, ..., n \quad (7)$$

$$1 - (n+1)(1 - x_{ij}) \leq u_i - u_j \leq$$
$$\leq 1 + (n+1)(1 - x_{ij}), \; (i,j) \in E \quad (8)$$

$$t_j - t_i \leq -1 + (n+1)(2 - x_{ik} - x_{jk}) + (n+1)(1 - y_{ij})$$
$$i, j, k = 1, ..., n, \; i < j \quad (9)$$

$$1 - (n+1) y_{ij} \leq t_j - t_i \leq (n+1)(1 - y_{ij}),$$
$$i, j = 1, ..., n, \; i < j \quad (10)$$

$$t_i + 1 - (n+1)(1 - x_{ij}) \leq t_j, \; (i,j) \in E \quad (11)$$

Constraints (7) guarantee that each vertex sends a message only once during the aggregation session. Constraints (8) ensure that the subgraph which is defined by the variables **x** is a tree. With the constraints (9) and (10) the conflicts between children of a same parent are eliminated. The constraints (11) hold the requirement that each vertex can transmit data only after receiving messages from all of its children in the aggregation tree.

### 3.2 Comparison of the IP formulations

We have tested the both IP formulations using the IBM ILOG CPLEX package. We launched CPLEX on commonly used interconnection topologies: butterfly graph ($BF_d$), cube connected cycle ($CCC_d$) and shuffle-exchange graph ($SE_d$) (Table 1). More detailed information about these graph classes can be found in [25]. We also launched CPLEX for instances generated randomly using GT-ITM Pure Random model [26]. Results of the experiments are presented in Table 1 and Table 2, respectively. $CPLEX_{IP1}$ stands for the CPLEX using formulation (1)-(4) and $CPLEX_{IP2}$ stands for the CPLEX using formulation (5)-(11). The calculation time was limited by 1000 seconds. If CPLEX failed to find an optimal solution during 1000 seconds, then the best found feasible solution was returned. In this case the objective value is marked in italics.

The results of the experiment show that the both IP formulations are suitable to solve the considered problem in acceptable time in cases of small dimension (10-25 vertices and 20-50 edges). When $n \geq 40$ and $m \geq 60$ $CPLEX_{IP2}$ is unable to complete the process during 1000 seconds (except one case when $n = 40$, $m = 64$), but it always finds an optimal or near-optimal solution. This means that $CPLEX_{IP2}$ finds a near-optimal solution rather fast, and spends the majority of running time for the proof of its optimality. Although CPLEX$_{IP1}$ appeared to outperform CPLEX$_{IP2}$ in some cases, the results of CPLEX$_{IP1}$ were significantly worse when the calculation process was aborted due the time limit. Additionally, CPLEX$_{IP1}$ was often unable to find any feasible solution (see, e.g. cases when $n \geq 50$ in Table 2). In summary, we conclude that the IP formulation (5)-(11) is more preferable in practice to solve the considered problem by CPLEX than the formulation (1)-(4). Even if CPLEX fails to find an optimal solution for the formulation (5)-(11) in a specified time, it always provides a decent near-optimal feasible solution.

**Table 1.** CPLEX performances on the popular network topologies.

| | $n$ | $m$ | CPLEX$_{IP1}$ | | CPLEX$_{IP2}$ | |
|---|---|---|---|---|---|---|
| | | | Time (sec.) | Obj | Time (sec.) | Obj |
| CCC$_3$ | 24 | 36 | 8.92 | 6 | 4.99 | 6 |
| CCC$_4$ | 64 | 96 | 1000 | *60* | 1000 | *10* |
| SE$_3$ | 8 | 10 | 0.02 | 5 | 0.01 | 5 |
| SE$_4$ | 16 | 21 | 0.67 | 7 | 0.19 | 7 |
| SE$_5$ | 32 | 46 | 44.73 | 9 | 159.1 | 9 |
| BF$_3$ | 24 | 48 | 96.44 | 5 | 87.49 | 5 |
| BF$_4$ | 64 | 128 | 1000 | *58* | 1000 | *8* |

**Table 2.** CPLEX performances on the pure random graphs.

| $n$ | $m$ | CPLEX$_{IP1}$ | | CPLEX$_{IP2}$ | |
|---|---|---|---|---|---|
| | | Time (sec.) | Obj | Time (sec.) | Obj |
| 10 | 29 | 2.329 | 4 | 2.16 | 4 |
| 10 | 26 | 0.703 | 4 | 0.856 | 4 |
| 25 | 35 | 3.96 | 5 | 0.85 | 5 |
| 25 | 40 | 8.29 | 6 | 10.82 | 6 |
| 25 | 47 | 10.99 | 5 | 119.1 | 5 |
| 40 | 64 | 1000 | *30* | 965 | 7 |
| 40 | 68 | 67,6 | 7 | 1000 | 7 |
| 40 | 70 | 46,4 | 6 | 1000 | 6 |
| 50 | 158 | 1000 | *42* | 1000 | 7 |
| 50 | 127 | 1000 | - | 1000 | *6* |
| 50 | 126 | 1000 | - | 1000 | *7* |
| 100 | 232 | 1000 | - | 1000 | *9* |
| 100 | 233 | 1000 | - | 1000 | *10* |
| 100 | 367 | 1000 | - | 1000 | *9* |

## 4 Heuristic algorithm GLS

In this section, we propose the new heuristic algorithm, Genetic Local Search (GLS), which combines a genetic algorithm approach [27] with a local search meta-heuristic. Similar to a conventional genetic algorithm, GLS maintains a set of feasible solutions (*population*) and imitates an evolutionary process as follows: at each iteration the pairs of solutions are chosen from the population and reproduce an *offspring*. As soon as a new solution is generated, it can be modified by the Mutation procedure. After this, the Local Search procedure tries to improve the current solution. Each time the best solutions are kept in the population of the next generation. This process continues until some predefined stopping condition is met.

The pseudocode of the GLS algorithm is presented in Fig. 1. The starting population is generated at the *Initialization* step in line 1. After that in lines 3-9 the following steps are sequentially repeated until a stopping condition is met: *Selection*, *Crossover*, *Mutation*, *LocalSearch*, *FitnessCalculation* and *Join*.

As an input the algorithm takes a communication graph $G_{or}$ and the following set of parameters:
- *PopSize* – the size of population;
- *OffspSize* – the size of offspring;
- *FPItCount* – the number of iterations in the first population construction procedure;
- *SPProportion* – the ratio of shortest-path trees in the starting population;
- *PM* – the probability of mutation;
- *PLS* – the probability of local search.
- $k_{max}$ – the maximum possible number of iterations in mutation procedure

The next subsections contain detailed descriptions of the algorithm steps.

### 4.1 Initialization

At the *Initialization* step the first population is generated. The first tree, which is added into the first population, is the shortest-path tree constructed by the Dijkstra algorithm. After this tree is constructed, the length of the shortest path from each vertex to the sink is known. Let $l(v)$ be the length (number of edges) of a shortest path from vertex $v \in V$. Let us consider a directed graph $G_1 = (V, A_1)$, where $A_1 = \{(u,v) \mid (u,v) \in A, \ l(u) = l(v) - 1\}$. Note that any spanning tree which is rooted in $s$ and contains only arcs from $A_1$ is a shortest-path tree. The next trees added to the population are generated by two procedures: *RandomShortestPath* and *RandomMinDegree*. The procedure *RandomShortestPath*

```
    INPUT: G_or = (V, A) - communication graph, PopSize,
OffspSize, FPItCoun, SPProportion, PM, PLS, k_max -
additional parameters;
    OUTPUT: T - spanning tree on G rooted in s;
  1.  Initialization;
  2.  FitnessCalculation(population);
  3.  while (stop condition is not met)
  4.     Selection;
  5.     Crossover;
  6.     Mutation;
  7.     LocalSearch;
  8.     FitnessCalculation(offspring);
  9.     Join;
  10.    T = the best tree among the current population
  11. end while
```
**Fig. 1.** Genetic local search (GLS).

starts with a tree $T = (\emptyset, \{s\})$; an arc from $A_1$ which connects a vertex from the current tree with a vertex from $V$ which is not in the current tree is sequentially chosen at random and added to the current tree. In the procedure *RandomMinDegree*, the tree is constructed in a similar manner, but with the following difference: at each step an arc is chosen randomly from $A$, and the probability of an arc choice is inversely proportional to the degree of a corresponding vertex in the current tree. A new tree is added to the population only if it is not a copy of an existing one. The *Initialization* step requires three parameters: *PopSize* – the maximum size of the population, *SPProportion* – an approximate part of the trees generated by the procedure *RandomShortestPath*, and *FPItCount* – the maximum number of successive attempts to generate a tree. The pseudocode of the *Initialization* step can be found in Fig. 2.

```
    INPUT: G_or = (V, A) - communication graph,
SPProportion, FPItCount, PopSize - additional
parameters;
    OUTPUT: p - population (a set of spanning trees on G
rooted in s);
  1.  i = 0;
  2.  T₀ = Dijkstra(); // Dijkstra algorithm
  3.  p = {T₀}; // population
  4.  while (i < FPItCount and p.Size < PopSize)
  5.     p = random real value between 0 and 1
  6.     if (p < SPProportion)
  7.        T = RandomShortestPath();
  8.     else
  9.        T = RandomMinDegree();
  10.    if (p contains T)
  11.       i++; // clones are forbidden
  12.    else
  13.       p.add(T);
  14. end while
```
**Fig. 2.** GLS: Initialization.

## 4.2 Fitness calculation

In order to estimate the quality of every solution in the population its *fitness* should be calculated. Fitness is a positive value which is higher when the solution is closer to the optimal solution. Let $L(T)$ be the minimum convergecasting schedule length for a spanning tree $T$. Then the fitness is $1 / L(T)$.

Note that the convergecasting schedule of minimum length on a spanning tree $T$ can be found in time $O(n)$ for example using the procedure described in [8] with a small modification, because the position of the broadcast center is known in our case.

## 4.3 Selection

In the Selection step a set of parents is filled by the solutions from the current population in the following way. Sequentially a tree is taken from the current population with proportion to fitness probability. Note that the same solution can be added to the parent set several times. The number of elements in the parent set exceeds twice the maximum number of elements in offspring *OffspSize*, which is the parameter of GLS.

## 4.4 Crossover

At first a set of parents is divided randomly into *OffspSize* pairs. After that each pair of parents $T_p^1 = (V, A_p^1)$ and $T_p^2 = (V, A_p^2)$ generates a child tree $T_c$ in the following way. Let us consider a vertex $v \in V \setminus \{s\}$ and two vertices $v_1, v_2 \in V$: $a_1 = (v, v_1) \in A_p^1$, $a_2 = (v, v_2) \in A_p^2$. The goal is to choose an arc from $\{a_1, a_2\}$ and to add it to $T_c$. If $v_1 = v_2$ then the arc $a_1$ is chosen. If adding of one arc from $\{a_1, a_2\}$ to $T_c$ leads to the appearance of cycles, then another arc is chosen. In the remaining case let us introduce the weight $w_i = 1 / \delta(v_i) + 1 / |l(v) - l(v_i) - 2|$, where $\delta(v_i)$ is a degree of the vertex $v_i$ in the tree $T_p^i$, $i \in \{1, 2\}$. Then the arc is chosen randomly from $\{a_1, a_2\}$ with probability $P(a_i) = w_i / (w_1 + w_2)$, $i \in \{1, 2\}$.

## 4.5 Mutation

Mutation is a randomized procedure which is applied to the solutions in the current offspring. The Mutation procedure is applied with probability $P_M$ (a parameter of GLS) to each offspring. The Mutation procedure takes as an argument (an integer parameter) $k$ – the maximum difference (number of different arcs) between the initial tree and the modified one. This parameter is taken randomly from the interval $[0, ..., k_{max}]$, where $k_{max}$ is another algorithm parameter, inverse to its value probability (i.e., smaller modifications are more possible). The pseudocode of the mutation procedure is given in Fig. 3.

```
    INPUT: G = (V, A) - communication graph, T = (V,
A(T)) - spanning tree on G rooted in s, k - an integer
parameter;
    OUTPUT: T - spanning tree on G rooted in s;
    1.  do k times:
    2.     (i,j) = random arc from A \ A(T);
    3.     if (j is not descendant of i)
    4.        T.SetParent(i, j);
```
**Fig. 3.** GLS: Mutation.

### 4.5 Local search

As well as mutation, the local search procedure is applied to a subset of offspring defined by the probability $P_{LS}$ – another algorithm parameter. The pseudocode of the local search procedure is presented in Fig. 4. At each iteration the procedure performs a search of such arc $a = (v_1, v_2) \in A \setminus A(T)$ whose addition of $T$ (together with detaching of $v_1$ from its parent in $T$) leads to the maximum decrease of the objective function. The method *CalculateEffect*($T$, $u$, $v$) calculates the change of the schedule length after detaching of $v$ from its parent in $T$ and adding an arc $(v, u)$. Method *T.SetParent* performs the mentioned reattaching. The whole procedure continues while the solution is improved.

```
    INPUT: G = (V, A) - communication graph, T = (V,
A(T)) - spanning tree on G rooted in s, k - an integer
parameter;
    OUTPUT: T - spanning tree on G rooted in s;
improved = true;
    1.  while (improved)
    2.     improved = false;
    3.     best_u = NULL; best_v = NULL; bestImpr = 0;
    4.     for each arc (u, v) in A \ A(T)
    5.        effect = CalculateEffect(T, u, v);
    6.        if (effect < bestImpr)
    7.           best_u = u;
    8.           best_v = v;
    9.           bestImpr = effect;
    10.          improved = true;
    11.       end if
    12.    end for
    13.    if (improved)
    14.       T.SetParent(best_u, best_v);
    15.       CalculateSchedule(T);
    16.       break;
    17.    end if
    18. end while
```
**Fig. 4.** GLS: Local search.

### 4.6 Join

At the join step *PopSize* solutions from the current population and the current offspring, which have the highest fitness, are chosen to fill the population of the next generation.

## 5 Simulation

The proposed algorithm has been implemented in C++. Also, we implemented two of the most efficient previous heuristics: the Round Heuristic (RH) [22] and the Tree Based Algorithm (TBA) [23] in order to compare them with our GLS algorithm. As test instances, the following commonly used interconnection topologies have been considered: $BF_d$ – butterfly graph, $CCC_d$ – cube connected cycle, $SE_d$ – shuffle-exchange graph. The results can be found in Tables 3-5. In addition, we ran all algorithms on random graphs generated using three well-known network models: GT-ITM Pure Random, GT-ITM Transit-Stub [26] and BRITE Top-Down Waxman [28], see Tables 6-8. OPT stands for the optimal value of the objective either known previously [25] or obtained by us using CPLEX launched on one of the IP formulations proposed in Section 3 (as for GT-ITM Pure Random model when n ≤ 25). If the optimal value of the objective is not known, then the lower bound (LB) is mentioned. For graphs $CCC_d$ and $BF_d$ the best known lower bounds are taken from [25]. For other models the maximum of the following two values are taken: (a) $\lceil \log_2 n \rceil$ [25]; (b) a value obtained by calculating for each vertex its minimum receiving time plus the length of the shortest path to the root. The minimal values of convergecasting time are marked bold in the tables below.

The following algorithm's parameters allow us to get the best results: *PopSize* = 50, *OffspSize* = 25, *FPItCount* = 3, *SPProportion* = 0.6, *PM* = 0.6, *PLS* = 0.8, $k_{max} = \lfloor n/3 \rfloor$. As a stopping criterion we used the following rule: the minimum and the maximum values of fitness among all solutions in the current population are not changed during last 10 iterations.

For some reason in several cases we failed to reproduce the results of TBA presented in [23]. For example, in $CCC_6$, $CCC_8$, $BF_6$, $BF_7$, $BF_8$, $SE_8$, the length of the schedule yielded by TBA in our implementation appeared to be greater by 1 than the length of the schedule constructed by RH, although the authors of TBA state that they are the same in those cases.

**Table 3.** Results in $CCC_d$.

| d | n | m | LB | RH | TBA | GLS |
|---|---|---|---|---|---|---|
| 3 | 8 | 12 | 6 | **6** | **6** | **6** |
| 4 | 24 | 36 | 9 | **9** | **9** | **9** |
| 5 | 160 | 240 | 11 | **11** | **11** | **11** |
| 6 | 384 | 576 | 13 | **13** | 14 | **13** |
| 7 | 896 | 1344 | 16 | **16** | **16** | **16** |
| 8 | 2048 | 3072 | 18 | **18** | 19 | **18** |

**Table 4.** Results in $BF_d$.

| d | n | m | LB | RH | TBA | GLS |
|---|---|---|---|---|---|---|
| 3 | 24 | 48 | 5 | **5** | **5** | **5** |
| 4 | 64 | 128 | 7 | **7** | **7** | **7** |
| 5 | 160 | 320 | 8 | **9** | **9** | **9** |

| 6 | 384 | 768 | 10 | **10** | 11 | **10** |
| 7 | 896 | 1792 | 11 | **12** | 13 | **12** |
| 8 | 2048 | 4096 | 13 | **14** | 15 | **14** |

Table 5. Results in $SE_d$.

| d | n | m | OPT | RH | TBA | GLS |
|---|---|---|-----|----|----|-----|
| 3 | 8 | 10 | 5 | **5** | **5** | **5** |
| 4 | 16 | 21 | 7 | **7** | **7** | **7** |
| 5 | 32 | 46 | 9 | **9** | **9** | **9** |
| 6 | 64 | 93 | 11 | **11** | **11** | **11** |
| 7 | 128 | 190 | 13 | **13** | **13** | **13** |
| 8 | 256 | 381 | 15 | **15** | 16 | **15** |

Table 6. Results in GT-ITM Pure Random model.

| n | m | LB | OPT | RH | TBA | GLS |
|---|---|----|-----|----|----|-----|
| 10 | 29 | 4 | 4 | **4** | **4** | **4** |
| 10 | 23 | 4 | 4 | **4** | **4** | **4** |
| 25 | 35 | 5 | 5 | 6 | 6 | **5** |
| 25 | 40 | 6 | 6 | **6** | 7 | **6** |
| 25 | 47 | 5 | 5 | **5** | **5** | **5** |
| 25 | 42 | 5 | 6 | **6** | **6** | **6** |
| 100 | 277 | 7 | - | **7** | **7** | **7** |
| 100 | 232 | 7 | - | 8 | 8 | **7** |
| 100 | 233 | 7 | - | **8** | 9 | **8** |
| 500 | 1497 | 9 | - | **10** | **10** | **10** |
| 500 | 1259 | 9 | - | **10** | **10** | **10** |
| 500 | 1529 | 9 | | **11** | 12 | **11** |
| 500 | 1529 | 9 | | **12** | **12** | **12** |
| 500 | 1530 | 10 | | **12** | 13 | **12** |
| 1000 | 1028 | 22 | | **22** | **22** | **22** |
| 1000 | 2020 | 17 | | **19** | **19** | **19** |
| 1000 | 3027 | 10 | | **13** | 14 | **13** |

Table 7. Results in BRITE Top-Down Waxman model.

| n | m | LB | RH | TBA | GLS |
|---|---|----|----|----|-----|
| 100 | 208 | 8 | **10** | 11 | **10** |
| 100 | 307 | 9 | 11 | **10** | **10** |
| 100 | 407 | 9 | **10** | **10** | **10** |

Table 8. Results in GT-ITM Transit-Stub model.

| n | m | LB | RH | TBA | GLS |
|---|---|----|----|----|-----|
| 100 | 287 | 7 | **8** | 9 | **8** |
| 100 | 261 | 7 | **9** | **9** | **9** |
| 100 | 267 | 7 | **9** | 11 | **9** |
| 100 | 273 | 7 | 9 | 9 | **8** |
| 100 | 275 | 7 | 9 | **8** | 9 |
| 600 | 1004 | 11 | **14** | **14** | **14** |
| 600 | 1208 | 10 | 14 | **13** | **13** |
| 600 | 1250 | 10 | 15 | 14 | **13** |
| 600 | 1234 | 10 | 15 | **14** | **14** |
| 600 | 1235 | 10 | 13 | 13 | **12** |
| 1020 | 2533 | 10 | **16** | **16** | **16** |
| 1020 | 3366 | 10 | 17 | 17 | **16** |
| 1020 | 3515 | 10 | 17 | **16** | **16** |
| 1020 | 2563 | 10 | 18 | **17** | **17** |
| 1020 | 2550 | 10 | 17 | 17 | **16** |

The results of the experiment show that our algorithm GLS performs better or the same as other known heuristics. The most noticeable advantage of GLS can be observed on the GT-ITM Transit-Stub model: in 10 cases out of 15 it outperforms at least one of the other algorithms, in 4 cases it outperforms both of them, and in 2 cases its convergecasting time is less by 2 than the convergecasting time obtained by one of other algorithms. We found only one case when another algorithm outperforms GLS: in the fifth instance of the BRITE Transit-Stub model the aggregation latency obtained by TBA is less by 1 than the one obtained by

GLS. In total, GLS appeared to be extremely efficient for all tested topologies and network models.

Obviously, GLS is more time-consuming than RH and TBA, but its running time remained acceptable in all tested cases. Thus, it solves a problem with 100 vertices and 300 edges in about 1.5 seconds, and it solves a problem with 1000 vertices and 3000 edges in about 30 seconds. It should be noticed that we launched it on one thread although it is well-parallelizable and therefore it may be significantly speeded-up according to the machine hardware and operation system properties. Moreover, the variety of parameters of GLS provides flexibility, and more thorough tuning of these parameters may also improve the algorithm.

# 6 Conclusion

In this paper, we addressed an aggregated convergecast problem in WSNs for a case when the number of channels is unbounded. The objective is to minimize the data aggregation time. We proposed a new heuristic algorithm, which is based on the genetic algorithm and the local search metaheuristic. Our scheduling algorithm has lower latencies than the previous best approaches, Round Heuristic [22] and Tree Based Algorithm [23] especially for such widespread network model as GT-ITM Transit–Stub. We also proposed a new IP-formulation to this problem and compared it with a variant from [9] using IBM CPLEX package. Our formulation appeared to be more suitable for moderate-size instances (up to 100 vertices and 300 edges), where it always finds a near-optimal feasible solution in a reasonable time.

The research of R. Plotnikov is partly supported by the Russian Foundation for Basic Research (grant no. 16-37-60006). The research of A. Erzin and V. Zalyubovskiy is partly supported by the Russian Foundation for Basic Research (grant no. 16-07-00552).